\documentclass[journal=nalefd,manuscript=letter]{achemso}

\usepackage{amssymb}
\usepackage{color}

\newcommand*\ICN{Catalan Institute of Nanoscience and Nanotechnology (ICN2), CSIC and The Barcelona Institute of Science and Technology (BIST), Campus UAB, Bellaterra, 08193 Barcelona, Spain}
\newcommand*\OSAKA{Graduate School of Engineering Science and Center for Spintronics Research Network (CSRN), Osaka University, Toyonaka, Osaka 560-8531, Japan}
\newcommand*\SPINTEC{SPINTEC, Univ. Grenoble Alpes, CNRS, CEA, 17 rue des Martyrs, 38054 Grenoble, France}
\newcommand*\ICREA{Instituci\'{o} Catalana de Recerca i Estudis Avan\c{c}ats (ICREA), 08070 Barcelona, Spain}

\author{Fr\'{e}d\'{e}ric Bonell}
\affiliation{\ICN} \altaffiliation{Current address: \SPINTEC} \email{frederic.bonell@cea.fr}
\author{Minori Goto}
\affiliation{\OSAKA}
\author{Guillaume Sauthier} \author{Juan F. Sierra} \author{Adriana I. Figueroa} \author{Marius V. Costache}
\affiliation{\ICN}
\author{Shinji Miwa} \author{Yoshishige Suzuki}
\affiliation{\OSAKA}
\author{Sergio O. Valenzuela}
\affiliation{\ICN} \alsoaffiliation{\ICREA} \email{SOV@icrea.cat}

\title{Control of spin-orbit torques by interface engineering in topological insulator heterostructures}


\begin{document}

\begin{abstract}
(Bi$_{1-x}$Sb$_x$)$_2$Te$_3$ topological insulators (TIs) are gathering increasing attention owing to their large charge-to-spin conversion efficiency and the ensuing spin-orbit torques (SOTs) that can be used to manipulate the magnetization of a ferromagnet (FM). The origin of the torques, however, remains elusive, while the implications of hybridized states and the strong material intermixing at the TI/FM interface are essentially unexplored. By combining interface chemical analysis and spin-transfer ferromagnetic resonance (ST-FMR) measurements, we demonstrate that intermixing plays a critical role in the generation of SOTs. By inserting a suitable normal metal spacer, material intermixing is reduced and the TI properties at the interface are largely improved, resulting in strong variations in the nature of the SOTs. A dramatic enhancement of a field-like torque, opposing and surpassing the Oersted-field torque, is observed, which can be attributed to the non-equilibrium spin density in Rashba-split surface bands and to the suppression of spin memory loss.
\end{abstract}

\newpage
The search for efficient SOTs using TIs has been motivated by their topological boundary states characterized by spin-momentum locking \cite{Hasan2010,Qi2011}. There, an electron spin orientation is fixed relative to its propagation direction (Fig. 1a). Thus, when a current flows at the boundary of a TI a significant non-equilibrium spin density is expected. This effect is equivalent to the Rashba-Edelstein effect (REE) in two-dimensional electron gases (2DEGs) \cite{Edelstein1990} (also known as inverse spin galvanic effect). The non-equilibrium spins could then exert torques onto the magnetization of a neighboring FM, manipulating its dynamics or even helping switch its orientation \cite{Miron2010,Mahendra2018,Khang2018}. Recent observations of anomalously large torques in TI/FM bilayers have been attributed to topological boundary states \cite{Mellnik2014,Wang2015,Kondou2016,Wang2017,Mahendra2018,Khang2018}. However, the experiments have been mainly performed with highly doped TIs (Bi$_2$Se$_3$, Bi$_2$Te$_3$, and Sb$_2$Te$_3$) in which the topological states coexist with bulk bands at the Fermi level and contribute only marginally to the electrical conduction, especially at room temperature \cite{Costache2014}. Furthermore, the analysis of the results follows the implicit assumption that the TI/FM interface is perfectly abrupt, despite existing reports demonstrating the presence of magnetic dead layers \cite{Li2012,Wang2015}, strong intermetallic alloying \cite{Li2012,Walsh2017,Ferfolja2018}, and band bending \cite{King2011,Bahramy2012,Valla2012}. Alternative torque mechanisms were disregarded under such an assumption. The estimated torques due to the bulk spin Hall effect (SHE) and current-induced Oersted fields were deemed too small to explain the experimental observations \cite{Mellnik2014}. It has been further argued that spin density from Rashba-split 2DEGs, originating from surface bending of bulk bands, would lead to a torque with opposite orientation to that observed \cite{Mellnik2014,Wang2015}. However, any deviation from an ideal interface could create, for instance, an additional conducting layer, which would shunt the current and inadvertently enhance the Oersted-field torque. Therefore, the already established strong material intermixing, leading to alloying and, possibly, magnetic dead layers, demands a clarification of interface effects in the interpretation of the observed torques. This is critical for the optimization of the SOTs and the application of TIs in magnetic memory technologies.

In order to reveal the nature of the torques when the interface properties change, we implement systematic studies in TI/FM and TI/NM/FM heterostructures, with selected normal metal (NM) spacers. We then analyze the results by considering the corresponding interface chemistry, as inferred from x-ray photoelectron spectroscopy (XPS). We show that the spacers, with variable thickness $t_{NM}$, modify the degree of intermixing and that, when the intermixing is suppressed, a field-like torque, that opposes the Oersted-field torque, is dramatically enhanced. Such findings are attributed to the non-equilibrium spin density associated to Rashba-split surface bands, which play a more significant role in establishing the torques than previously thought. The observed SOTs as a function of interface disorder are well accounted for by the spin memory loss mechanism applied to the non-equilibrium spin density.

\begin{figure}
\includegraphics[width=1\linewidth]{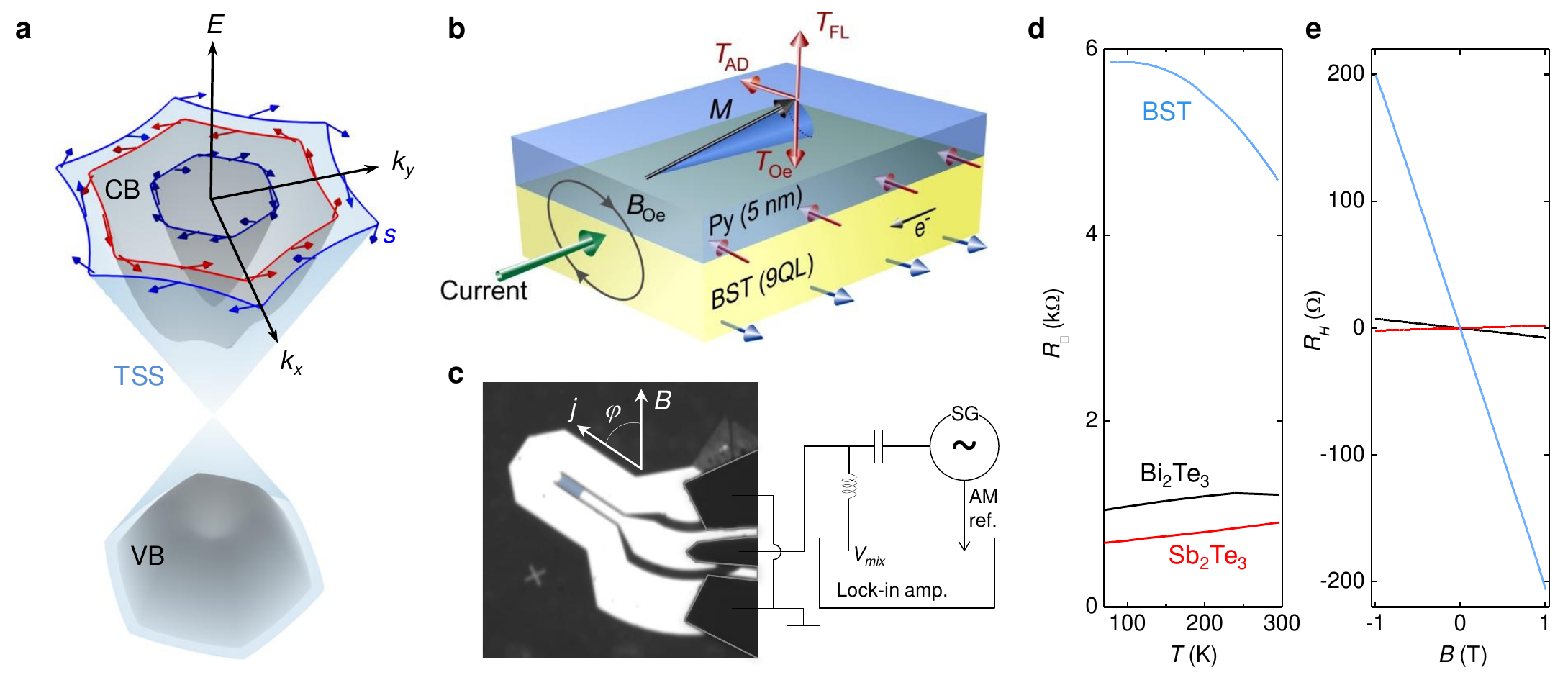}
\caption{Electronic properties of TIs and spin-orbit torques measurement scheme. (a) Representation of the TI electronic band structure (CB: conduction band; VB: valence band; TSS: topological surface states). The arrows represent the spin angular momentum. At TI/metal interfaces, a Rashba-split 2DEG generally develops in the CB due to strong band bending and quantum confinement. The outer Fermi contour of Rashba bands possesses a counterclockwise spin helicity, whereas the TSS possess a clockwise spin helicity \cite{Bahramy2012,King2011}. (b) Spin-orbit torques produced by the Rashba-Edelstein effect in a TI/FM bilayer. The arrows represent the non-equilibrium magnetization produced by an in-plane charge current. A counterclockwise spin helicity is expected to produce a field-like torque $T_{FL}$ opposed to the Oersted torque $T_{Oe}$. In contrast, the clockwise helicity of the topological surface states would produce a $T_{FL}$ that will add to $T_{Oe}$. (c) ST-FMR measurement scheme. An amplitude-modulated (AM) rf current is injected in the device from a signal generator (SG) through the rf port of a bias tee. The mixing voltage $V_{mix}$ is detected at the dc port by a lock-in method. (d) Sheet resistance of 9 nm-thick Bi$_{2}$Te$_3$, Sb$_{2}$Te$_3$ and (Bi$_{0.4}$Sb$_{0.6}$)$_2$Te$_3$ (BST) films capped with AlO$_y$(2 nm). (e) Hall resistance at 77K for the same films in (d).}
\label{Fig1}
\end{figure}

The heterostructures are fully grown in ultra-high vacuum (2$\times$10$^{-10}$ mbar) by molecular beam epitaxy on BaF$_2$(111) substrates, which favour intrinsically low-doped TIs with a very low density of structural defects \cite{Bonell2017}. They consist of BST(9nm)/NM($t_{NM}$)/Py($t_{FM}$) multilayers, where BST stands for (Bi$_{1-x}$Sb$_x$)$_2$Te$_3$ and Py for permalloy (Ni$_{0.8}$Fe$_{0.2}$); NM can be Al, Ag or Te. The growth of Al, Ag and Py is carried out at a substrate temperature of $\sim 120$ K in order to minimize intermixing at interfaces and to maximize the films coverage. All heterostructures are protected with a 2 nm-thick AlO$_y$ capping layer obtained by oxidation of an Al film in air.

Current-induced torques are characterized by means of ST-FMR (Fig. 1b), with the multilayers patterned by electron beam lithograhy and Ar ion etching into $100 \times 20$ $\mu$m$^2$ microbars and embedded in coplanar waveguides (Fig. 1c). In the ST-FMR experiments, we set $x = 0.6$ in order to minimize bulk conduction in the TI \cite{Kondou2016}. As seen in Fig. 1d, the sheet resistance $R_{\Box}$ is several times larger for (Bi$_{0.4}$Sb$_{0.6}$)$_2$Te$_3$ than for Bi$_2$Te$_3$ and Sb$_2$Te$_3$; it also increases at low temperatures, as expected for an insulator. Furthermore, the large $R_{\Box}$ is accompanied by an enhanced Hall resistance $R_H$ (Fig. 1e) that denotes a drop in the carrier density of about two orders of magnitude. The carrier density $n$ in (Bi$_{0.4}$Sb$_{0.6}$)$_2$Te$_3$, Bi$_2$Te$_3$ and Sb$_2$Te$_3$ at 77 K (room temperature) is respectively, $n=3\times10^{12}$ cm$^{-2}$ ($6.9\times10^{12}$ cm$^{-2}$), $n=8\times10^{13}$ cm$^{-2}$ ($1\times10^{14}$ cm$^{-2}$) and $n=3\times10^{14}$ cm$^{-2}$ ($2.9\times10^{14}$ cm$^{-2}$).

Figures 1b and 1c illustrate the experimental configuration for the ST-FMR measurements. Briefly, a microwave current $I_{rf}$ of fixed frequency $f$ (applied along the microbar) generates oscillatory torques. The FM is driven in and out-of the ferromagnetic resonance condition with an in-plane static magnetic field $\mathbf{B}$ at an angle $\varphi$ relative to the direction of the applied current (Fig. 1c). The dephasing between the current and the resistance oscillations, which derive from the anisotropic magnetoresistance of Py, results in a dc voltage $V_{mix}(B)$, which is measured with a lock-in amplifier and is used to characterize the SOTs.

The spin transfer mechanism and the magnetic exchange with non-equilibrium spins result in SOTs with distinct characteristics \cite{Wang2012,Fischer2016,Ndiaye2017,Haney2013}. The spin transfer mechanism leads to an in-plane (anti-)damping (AD) torque $T_{AD}$ that tends to align the magnetization with the current-induced magnetic moments (Fig. 1b). Direct exchange with the non-equilibrium magnetization $\delta \mathrm{\mathbf{m}}$, associated to the spin density, results in an out-of-plane field-like (FL) torque $T_{FL}$ that induces magnetization precession. As long as the FM is thicker than its transverse spin coherence length, the bulk SHE only contributes to $T_{AD}$ \cite{Haney2013} (in our devices $t_{FM} = 5$ nm, which is larger than the Py transverse spin coherence length: $1-2$ nm \cite{Ghosh2012}). Therefore, $T_{FL}$ can only originate from the topological surface states or the Rashba-split surface bands and is broadly considered to be their hallmark \cite{Wang2012,Fischer2016,Ndiaye2017,Haney2013}. For this reason, the discussion below concentrates on $T_{FL}$. However, because the torque $T_{Oe}$ produced by the current-induced Oersted field adds up to $T_{FL}$, a careful analysis and knowledge of the intermixing is required in order to properly discriminate it.

Figures 2 shows typical $V_{mix}$ for BST/Py as a function of $B$ (Fig. 2a), $\varphi$ (Fig. 2b, solid symbols), and power $P \propto I_{rf}^2$ (Fig. 2c, solid symbols). The torques $T_{AD}$ and $T_{Oe}+T_{FL}$, as defined in Fig. 1b, contribute, respectively, to the symmetric ($V_S$) and antisymmetric ($V_A$) Lorentzian components of the $V_{mix}(B)$ lineshape \cite{Liu2011,Mellnik2014,Wang2015,Kondou2016,Wang2017}, with $(T_{Oe}+T_{FL})/T_{AD} \propto V_A/V_S$ (see Supporting Information). Figures 2d presents ST-FMR results for representative NM spacers together with fits of the resonance lineshapes as the sum of symmetric and antisymmetric Lorentzians. All multilayers show a sizable $V_S$ with a sign that, in line with previous studies, coincides with that in Pt/Py \cite{Mellnik2014,Kondou2016}. In stark contrast, $V_A$ displays a remarkable dependence on the spacer characteristics. Indeed, for BST/Py, the sign of $V_A$ is the same as in Cu/Py and Pt/Py (Fig. 2e), which is equal to the sign of $T_{Oe}$ \cite{Mellnik2014}, however, for BST/Al/Py, the sign is opposite (see also open symbols in Figs. 2b and 2c and Supporting Information).

\begin{figure}
\includegraphics[width=1\linewidth]{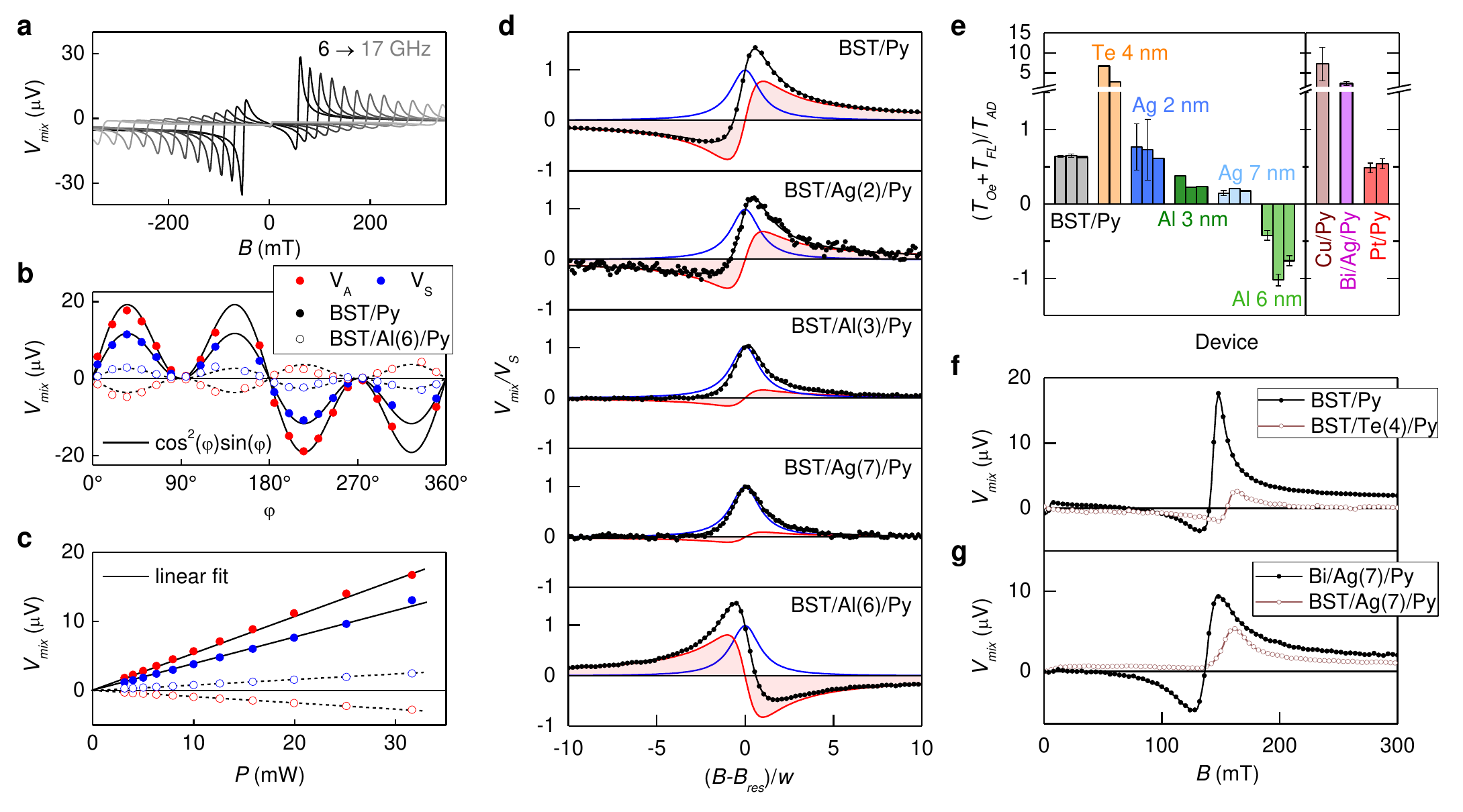}
\caption{ST-FMR measurements and interface engineering of the spin-orbit torques. (a) Typical ST-FMR spectra for a BST/Py device with excitation current $I_{rf}$ with frequency $f$ between 6 and 17 GHz. The spectra can be divided into symmetric ($V_S$) and antisymmetric ($V_S$) Lorentzians; their angular ($\varphi$) and power ($P$) dependencies are shown in (b) and (c), respectively for devices patterned on BST/Py (solid symbols) and BST/Al/Py (open symbols). (d) $V_{mix}$ for various NM spacer layers inserted at the BST/Py interface. The fitting curves (black) involve the sum of symmetric (blue) and antisymmetric (red) Lorentzians. In order to highlight the variation in $(T_{Oe}+T_{FL})/T_{AD}$, $V_{mix}$ is normalized to the amplitude of the symmetric component $V_{S}$ and plotted versus the reduced magnetic field ($B_{res}$ is the resonant field and $w$ the line width of the peaks). (e) $(T_{Oe}+T_{FL})/T_{AD}$ derived from $V_{A}/V_{S}$ for various devices and multilayers (see Supporting Information). Each bar corresponds to one device; the error bars represent the standard deviation in the range 5-15 GHz. (f) Effect of a Te spacer on ST-FMR spectra: due to strong spin absorption, Te fully suppresses the symmetric component. (g) Comparison of BST/Ag/Py and Bi/Ag/Py ST-FMR spectra. The Oersted torque dominates in the Bi/Ag/Py multilayer. In (a),(b),(d)-(g), $P$ = 15 dBm. In (b)-(d),(f),(g), $f$ = 9 GHz. In (a),(c)-(g), the magnetic field $B$ is applied at $\varphi=35^{\circ}$ from the current direction.}
 \label{Fig2}
\end{figure}

At first sight, the observed behavior is counterintuitive: adding a low-resistivity metallic spacer should increase the net Oersted field in the Py layer and lead to an enhanced $V_A/V_S$ ratio. In contradiction with this argument, Fig. 2d shows that $V_A/V_S$ decreases with a layer of 2 nm-thick Ag, drops to almost zero with 7 nm-thick Ag and 3-nm thick Al, and changes sign with 6 nm-thick Al. This trend is consistently found in all 17 tested devices, as demonstrated in Fig. 2e, which shows the torque ratio $(T_{Oe}+T_{FL})/T_{AD}$.

Since the Oersted field does not change its sign, the emergence of a negative torque ratio clearly demonstrates the presence of a $T_{FL}$ opposing $T_{Oe}$ that is largest in all three devices with a 6 nm-thick Al spacer (Fig. 2e). Independently of the actual magnitude of the SOTs, this finding provides unambiguous evidence of a large SOT produced by a non-equilibrium spin density at the BST interface. Indeed, the SHE contribution to $T_{FL}$ is expected to be negligible \cite{Haney2013}. In addition, a combination of spin pumping and spin-to-charge conversion would only contribute to $V_S$ \cite{Mellnik2014,Liu2011}, whose sign remains unchanged.

An independent confirmation of the spin origin of the torque is obtained from the drastic reduction of the ST-FMR signal that is observed when excess Te, a heavy element with high spin-orbit coupling, is deposited in the interfacial region. Figure 2f compares $V_{mix}(B)$ for BST/Te(4nm)/Py and BST/Py. The signal is much weaker in the former and exhibits a purely antisymmetric lineshape compatible with $T_{Oe}$. The absence of a discernable symmetric component not only shows that Te fully suppresses the transmission of non-equilibrium spins (a spin diffusion length in Te much shorter than 4 nm is inferred) but also rules out the presence of thermal effects, such as the spin Seebeck effect, which would contribute to $V_S$.

The fact that $T_{FL}$ and $T_{Oe}$ have opposite signs implies that, for a given current polarity, $\mathrm{\mathbf{\delta m}}$ induces an effective field $\mathrm{\mathbf{B_{FL}}} = \Delta_{ex} \mathrm{\mathbf{\delta m}}/M_S^2$ that is opposite to the Oersted field $\mathrm{\mathbf{B_{Oe}}}$ ($M_S$ is the Py magnetization). Considering ferromagnetic exchange in Py ($\Delta_{ex}>0$), $T_{FL}$ must originate from an electron Fermi contour with counterclockwise spin helicity. Since topological surface states are known to bear a clockwise spin helicity, the sign of $T_{FL}$ suggests that the torques in BST/Al/Py are due to interfacial Rashba-split bands (Fig. 1a). This is consistent with the strong band bending that is known to occur at TI/metal interfaces \cite{Bahramy2012,Valla2012}. Therefore $T_{FL}$ can be attributed to REE in the 2DEG that develops in the TI, above the conduction band minimum, with downward band bending \cite{King2011}.

We carried out XPS measurements in order to compare the observed trends in $(T_{Oe}+T_{FL})/T_{AD}$ with the degree of intermixing at the TI interface. Figure 3a shows representative XPS data that characterize the interface chemistry in the Bi$_2$Te$_3$/NM/Py and Bi$_2$Te$_3$/Py multilayers. The thickness of Py, $t_{FM} = 2$ nm, is chosen to be smaller than the XPS probing depth, while $x = 0$ is selected in order to maximize Bi emission lines (Sb is known to substitute Bi and reacts similarly to it). The measurements for Bi$_2$Te$_3$ and pure Bi and Te are shown as references. The bare Bi$_2$Te$_3$ surface was obtained by removing the Te capping of a Bi$_2$Te$_3$/Te bilayer, following the procedure reported in Ref.~\citenum{Bonell2017}. The Bi-4$f^{7/2}$ and Te-3$d^{5/2}$ emission lines for bare Bi$_2$Te$_3$ appear at specific binding energies (157.6 eV and 572.1 eV, respectively) that are different to those of pure Bi (156.9 eV) and pure Te (573.0 eV). The energy shifts for Bi and Te in Bi$_2$Te$_3$, relative to those of pure elements, are opposite because they originate from the charge transfer in the alloy, revealing the cationic (anionic) nature of Bi (Te).

\begin{figure}
\includegraphics[width=1\linewidth]{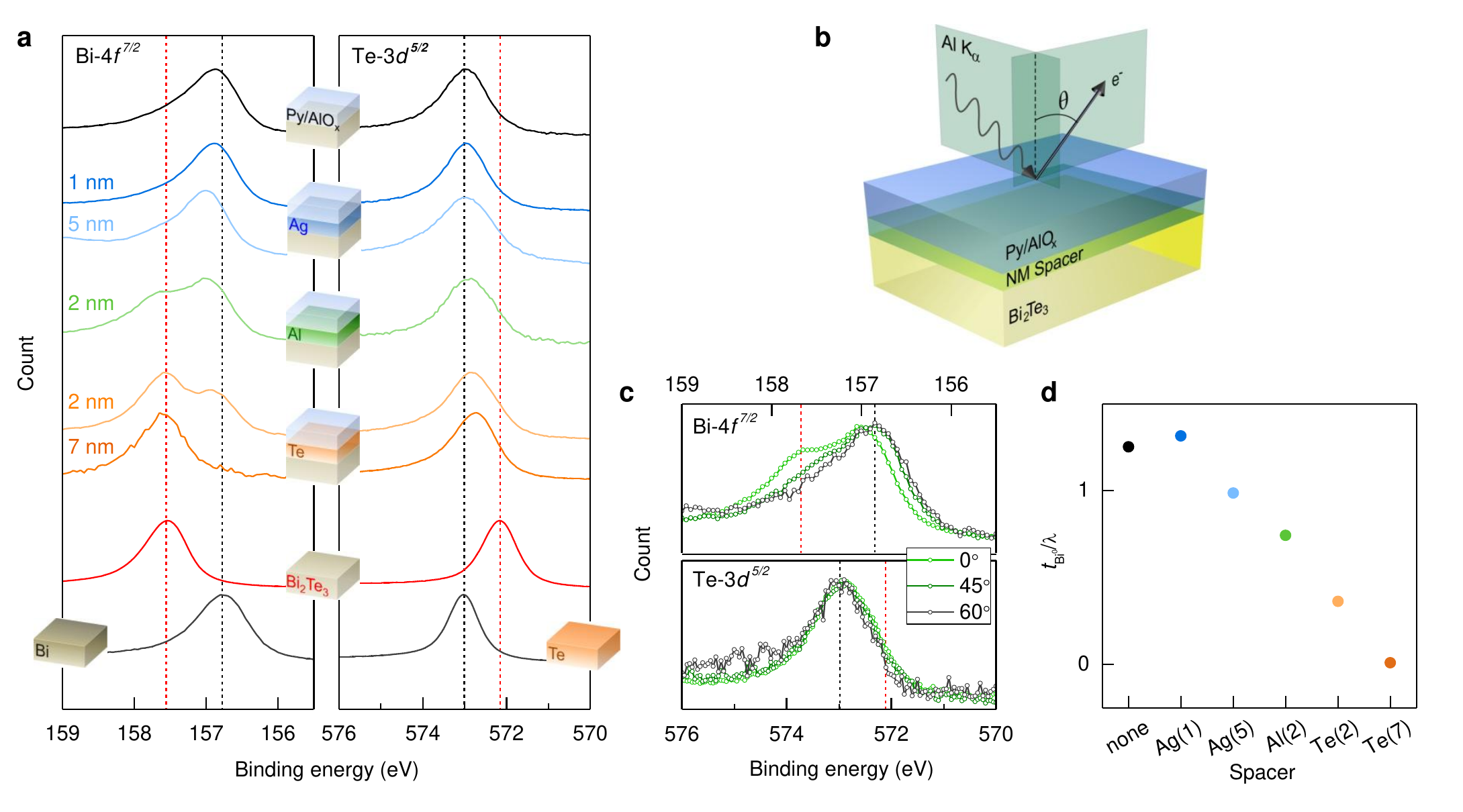}
\caption{Intermixing at TI/metal interfaces probed by XPS. (a) Bi-4$f^{7/2}$ and Te-3$d^{5/2}$ emission lines of Bi$_2$Te$_3$(9 nm)/NM/Py(2 nm)/AlO$_y$(2 nm) multilayers, where NM stands for Ag (1 nm or 5 nm), Al (2 nm) or Te (2 nm or 7 nm). Spectra of a bare Bi$_2$Te$_3$ surface are shown as references, as well as those of pure Te, from a Bi$_2$Te$_3$/Te(20 nm) stack, and pure Bi, from a Bi/Py(2 nm)/AlO$_y$(2 nm) stack. The characteristic energies for Bi-Bi, Te-Te and Bi-Te bonding are indicated by the vertical dashed lines. Bi-Bi and Te-Te bonding dominate the line shapes of Bi$_2$Te$_3$/Py. Characteristic Bi$_2$Te$_3$ line shapes are gradually recovered upon insertion of Ag, Al and Te spacer layers. (b) Measurement geometry. A larger detection angle $\theta$ ensures higher surface sensitivity. (c) Angular measurements with a 2 nm-thick Al spacer. (d) Quantification of the intermixed layer thickness $t_{Bi^0}$ from Bi-4$f^{7/2}$ line shapes. The thickness is given relative to the inelastic mean free path $\lambda_{Bi}$ in the intermixed layer ($\lambda_{Bi} \approx 2.5$ nm in a Bi-rich layer).}
\label{Fig3}
\end{figure}

Notably, the emission lines in Bi$_2$Te$_3$/Py are clearly distinct to those of bare Bi$_2$Te$_3$. The observed binding energies are actually characteristic of pure Bi and Te, which demonstrates that the Bi-Te covalent bonds are deeply altered due to the strong interaction with Py. Upon insertion of a few nanometer-thick Ag or Al layer between Bi$_2$Te$_3$ and Py, a second peak develops in the Bi-4$f^{7/2}$ emission line at 157.6 eV, which is the characteristic Bi binding energy in Bi$_2$Te$_3$. Although the two peaks are not resolved on Te-3$d^{5/2}$, a gradual broadening and red shift can be noticed. The presence of two components in the emission lines implies that Bi$_2$Te$_3$ and the intermixed interfacial layer coexist. However, the weight of the latter is significantly suppressed by the Ag and Al spacers.

Information on the location of the different chemical species and on the intermixing layer thickness is obtained with angular XPS. The escape depth of photoelectrons is given by $\lambda \cos \theta$, with $\lambda$ the photoelectron inelastic mean free path and $\theta$ the angle of detection from the surface normal. Therefore, the relative contribution of the surface is larger in grazing detection geometry (Fig. 3b). Angular XPS measurements for Bi$_2$Te$_3$/Al/Py are shown in Fig. 3c. It is observed that the Bi-4$f^{7/2}$ emission line exhibits a marked angular dependence that is absent in the Te-3$d^{5/2}$ emission line. Combined with the fact that the Bi-4$f^{7/2}$ peak corresponding to Bi$_2$Te$_3$ is strongest at normal detection, $\theta=0$, this implies that a Bi-rich region is buried under the Al/Py stack, while Te is more uniformly distributed in its volume. This suggests that Te diffuses out of Bi$_2$Te$_3$ into Al/Py, leaving a highly Te-depleted interface. Assuming two components in the Bi-4$f^{7/2}$ emission line, from Bi$_2$Te$_3$ and pure Bi, and taking $\lambda_{Bi}=2.5$ nm from Ref.~\citenum{Tanuma2011}, the effective thickness of the reduced Bi layer $t_{Bi^0}$ is estimated to be $t_{Bi^0} \approx 1.6$ nm (see Supporting Information for details). A similar analysis yields a significantly larger $t_{Bi^0} \approx 3$ nm at the direct Bi$_2$Te$_3$/Py interface, as expected from the dominant contributions of pure Bi and Te in the Bi$_2$Te$_3$/Py XPS spectra.

To further confirm the occurrence of Te out-diffusion, multilayers with an intentional excess of Te between Bi$_2$Te$_3$ and Py were fabricated. Excess Te is expected to diffuse in the metallic adlayers up to saturation, thereby protecting the Bi$_2$Te$_3$ surface. As seen in Fig. 3a, a Te spacer is indeed effective in limiting the Te out-diffusion from the Bi$_2$Te$_3$ film. Note that 7 nm-thick Te spacer has to be inserted so as to completely recover the characteristic Bi-4$f^{7/2}$ emission line of Bi$_2$Te$_3$, which confirms the large affinity of Py for Te. Figure 3d summarizes the extracted values of $t_{Bi^0}/\lambda_{Bi}$ for different spacers with their respective thickness. Besides Te, which is not transparent to spin transport, the smallest $t_{Bi^0}/\lambda_{Bi}$ is found for Al, which is the most efficient intermixing suppressor amongst the tested materials. Some intermixing suppression is also observed with increasing Ag thickness, providing an intermediate situation between the TI/Py and TI/Al interfaces.

Note that Rashba splitting is known to occur at Bi/NM interfaces and is particularly strong when the NM is Ag \cite{Ast2007}. In light of the XPS results (Fig. 3), reduced Bi is present at TI/NM interfaces, so that $T_{FL}$ might actually originate from it \cite{RojasSanchez2013}. However, ST-FMR measured in a control Bi(9nm)/Ag(7nm)/Py multilayer rules out this possibility (Fig. 2g). In this device, the resonance line shape is found to be almost purely antisymmetric, with the dominant contribution stemming from the Oersted field due to the charge current flowing in the Ag layer. Furthermore, from these measurements it can be estimated that the ratio $|T_{FL}/T_{Oe}|$ in BST/Ag/Py is about twice larger than in Bi/Ag/Py. This ratio is even larger for BST/Al/Py. Considering that the Bi/Ag bilayer exhibits one of the most efficient spin-to-charge conversion recorded \cite{RojasSanchez2013}, this result underscores the technological relevance of BST/Al bilayers.

The previous discussion provides compelling evidence that a gradual reduction of Te out-diffusion leads to a dramatic enhancement of $T_{FL}/T_{AD}$. Several spin-related phenomena are sensitive to the interface properties and can favor such an enhancement. In particular, the disorder deriving from intermixing and Te diffusion results in spin-current depolarization or spin memory loss (SML) \cite{RojasSanchez2014,Chen2015,Belashchenko2016,Dolui2017}. Furthermore, a spacer-dependent band bending could modulate the Rashba splitting and the Fermi energy in the interfacial 2DEG \cite{King2011}. When Te out-diffusion is suppressed, the Fermi level may approach the bottom of the Rashba-split bands where presumably the REE is largest. Complex spin textures arising from hybridization at the TI/NM interfaces may also impact the REE, although it is not obvious that they will result in $T_{FL}/T_{AD}$ variations \cite{MarmolejoTejada2017}.

\begin{figure}
\includegraphics[width=0.5\linewidth]{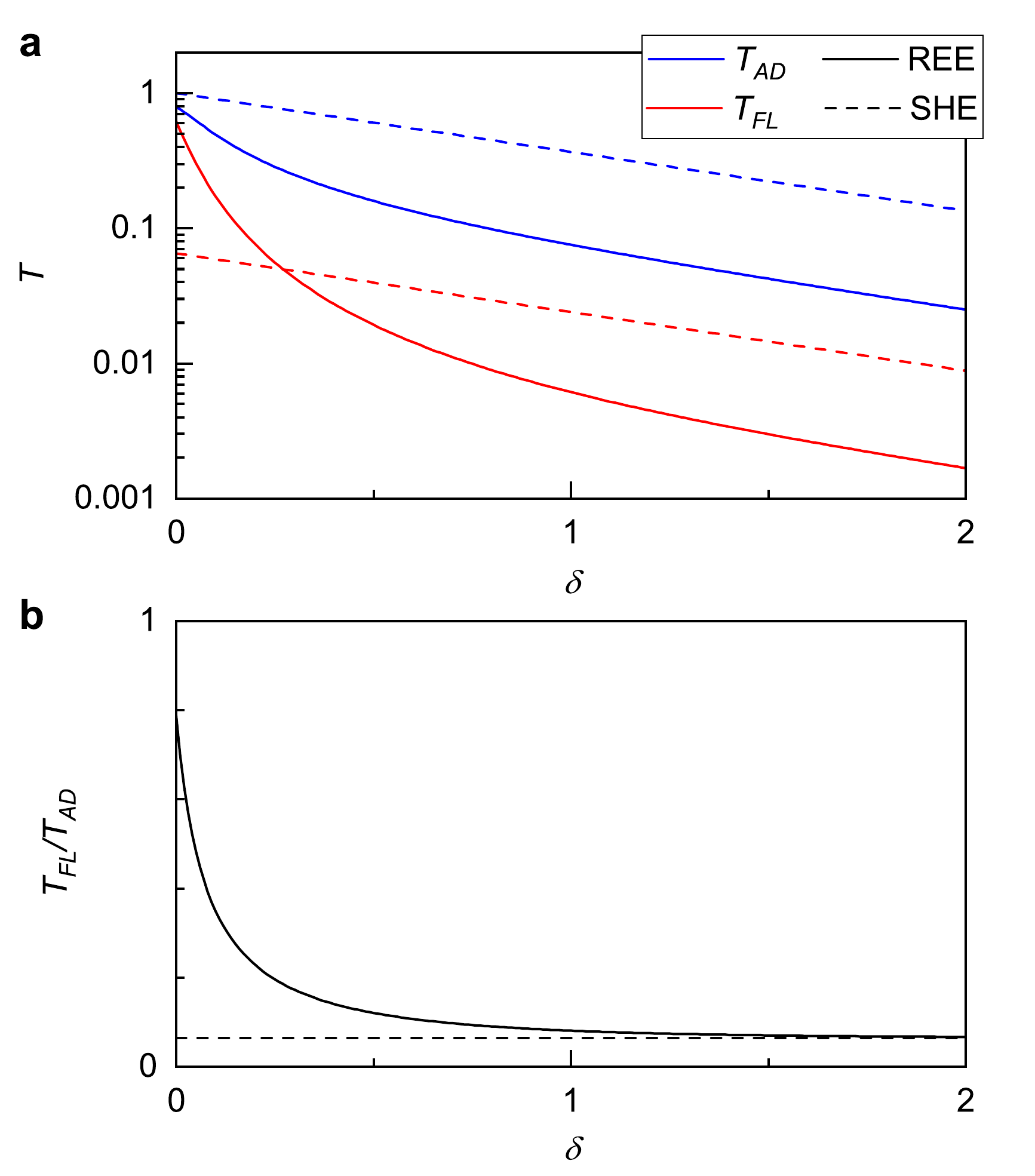}
\caption{Effect of spin memory loss on the spin-orbit torques. (a) Evolution of the field-like (FL) and anti-damping (AD) torques as a function of the spin memory loss parameter $\delta$. The decay of the torques is much faster for the Rashba-Edelstein effect (REE) than for the spin Hall effect (SHE). The torques are normalized by the magnitude of the total torque $|\mathbf{T}_{FL}+\mathbf{T}_{AD}|$. (b) Ratio of the torques. A large $T_{FL}$ is only expected with the REE and low spin memory loss ($\delta \ll 1$).}
 \label{Fig4}
\end{figure}

Despite the complexity of the involved phenomena, the experimental observations are actually fully captured by a simple model considering the SML. In order to estimate the effect of intermixing on $T_{FL}/T_{AD}$, we calculate $T_{FL}$ and $T_{AD}$ by solving the spin diffusion equation in a TI/I/FM trilayer, with I the interfacial region \cite{Fischer2016}. The considered SML is due to intermixing and disorder, which cannot be completely avoided with Ag or even Al, and leads to spin flips as the spins diffuse towards FM. Additional SML processes, deriving from interfacial spin-orbit coupling, could certainly be present but are not taken into consideration \cite{Chen2015,Belashchenko2016,Dolui2017}. The SML is quantified by the dimensionless parameter $\delta = t/\lambda_{sf}$, with $t$ the thickness of the interfacial region and $\lambda_{sf}$ its spin diffusion length. It is estimated that $\delta > 1$ with a Te spacer and $\delta < 1$ with the other interfaces. Figure 4 shows $T_{FL}$ and $T_{AD}$ (Fig. 4a) and $|T_{FL}/T_{AD}|$ (Fig. 4b) as a function of $\delta$ for torques induced by a bulk SHE (dashed lines) and by the REE (solid lines). The calculations assume a charge current oriented at $45^{\circ}$ from the FM magnetization, a uniform spin resistance through the structure, and thick TI and FM compared to their spin diffusion lengths (see Supporting Information). It is observed that in both the SHE and REE cases, the torques decrease with $\delta$, albeit much more rapidly for REE (Fig. 4a). Achieving abrupt interfaces thus enhances the torques but the enhancement is significantly more pronounced with a dominant REE. For the SHE, $T_{FL}$ is very small, as expected, and $|T_{FL}/T_{AD}|$ is independent of $\delta$ (Fig. 4b). In contrast, the REE can lead to $T_{FL}$ and $T_{AD}$ of similar magnitudes. In this case, a sizable $T_{FL}$ only exists when $\delta \ll 1$ and, as a consequence, $|T_{FL}/T_{AD}|$ is not constant but decreases with $\delta$. This implies that the increase in $|T_{FL}/T_{AD}|$ upon suppression of Te out-diffusion, as observed in the experiments (Fig. 2), is inconsistent with the SHE; instead, it points to a dominant REE and a reduced SML with Ag and Al spacers.

The decrease of $|T_{FL}/T_{AD}|$ with $\delta$ is intrinsic to the analysed SML process. The non-equilibrium spin density generated at the TI/I interface partially relaxes in the interfacial layer, turning into a net spin current towards the FM as in the SHE case. Eventually, for large $\delta$, the source of the spin current becomes indistinguishable and $|T_{FL}/T_{AD}|$ obtained for the REE converges to the value found for the SHE (Fig. 4b).

Our results demonstrate that deviations from an ideal TI/metal interface play a critical role in the generation of SOTs. Controlling the effect of such deviations provides a strategy to engineer TI/NM/Py heterostructures and to achieve efficient manipulation of magnetic devices. Intercalation of metallic spacers represents an effective way to reduce intermixing, and the spin memory loss, by preventing Te from diffusing out of the TI. Further engineering of the multilayers could ensure band-bending suppression and lower doping levels in the TI. This would help isolate the contribution of topological surface states, opening new perspectives to optimize the spin-orbit torques in topological insulator heterostructures. The mechanisms of Te diffusion and spin memory loss might play an important role at other chalcogenide interfaces of interest for efficient spin-charge interconversion, in particular those involving transition metal dichalcogenides, topological semimetals or two-dimensional FMs.

\begin{acknowledgement}
This research was partially supported by the European Research Council under Grant Agreement (GA) No. 306652 SPINBOUND, by the Spanish Ministry of Economy and Competitiveness, MINECO (under Contracts No MAT2016-75952-R and Severo Ochoa No. SEV-2017-0706), and by the CERCA Programme and the Secretariat for Universities and Research, Knowledge Department of the Generalitat de Catalunya 2017 SGR 827. F.B. acknowledges funding from the European Union's Marie Sk{\l}odowska-Curie actions under GA No. 624897 and from MINECO Ram\'{o}n y Cajal program under Contract No. RYC-2015-18523 and A.I.F. from the European Union's Marie Sk{\l}odowska-Curie actions under GA No. 796925.
\end{acknowledgement}

\begin{suppinfo}

Determination of the spin-orbit torques ratio from ST-FMR spectra, quantitative estimation of spin-orbit torque efficiencies, quantitative analysis of XPS spectra, analytic model of spin diffusion in a trilayer with spin memory loss.

\end{suppinfo}

\bibliography{Ref_STFMR_TIsInterlayers}

\end{document}